\documentclass[aps,pra,twocolumn,floatfix,superscriptaddress,showpacs]{revtex4}

\usepackage{graphicx}
\usepackage{bbm}
\usepackage{amsmath,amssymb,array}
\newcommand{\be}{\begin{equation}}
\newcommand{\beq}{\begin{equation}}
\newcommand{\ee}{\end{equation}}
\newcommand{\bea}{\begin{eqnarray}}
\newcommand{\eea}{\end{eqnarray}}
\newcommand{\ba}{\begin{array}}
\newcommand{\ea}{\end{array}}

\renewcommand{\vr} {{\bf r}}
\newcommand{\vj} {{\bf j}}
\newcommand{\vp} {{\bf p}}
\newcommand{\vs} {{\bf s}}

\begin{document}
\title{Gaussian approximations for the exchange-energy functional of current-carrying states:
Applications to two-dimensional systems}  
\author{S. Pittalis}
\email[Electronic address:\;]{pittalis@physik.fu-berlin.de}
\affiliation{Institut f{\"u}r Theoretische Physik,
Freie Universit{\"a}t Berlin, Arnimallee 14, D-14195 Berlin, Germany}
\affiliation{European Theoretical Spectroscopy Facility (ETSF)}
\author{E. R{\"a}s{\"a}nen}
\email[Electronic address:\;]{erasanen@jyu.fi}
\affiliation{Institut f{\"u}r Theoretische Physik,
Freie Universit{\"a}t Berlin, Arnimallee 14, D-14195 Berlin, Germany}
\affiliation{Nanoscience Center, Department of Physics, 
University of Jyv{\"a}skyl{\"a}, FI-40014 Jyv{\"a}skyl{\"a}, Finland}
\affiliation{European Theoretical Spectroscopy Facility (ETSF)}
\author{E. K. U. Gross}
\affiliation{Institut f{\"u}r Theoretische Physik,
Freie Universit{\"a}t Berlin, Arnimallee 14, D-14195 Berlin, Germany}
\affiliation{European Theoretical Spectroscopy Facility (ETSF)}
\affiliation{Max-Planck-Institut f{\"u}r Mikrostrukturphysik, 
Weinberg 2, D-06120 Halle, Germany}

\date{\today}

\begin{abstract}
Electronic structure calculations are routinely carried out within the 
framework of density-functional theory, often with great success.
For electrons in reduced dimensions, however, 
there is still a need for better approximations to the
exchange-correlation energy functional.
Furthermore, the need for properly describing
current-carrying states 
represents an additional challenge for the development of approximate
functionals. In order to make progress along these directions, 
we show that simple and efficient expressions for the exchange energy 
can be obtained by considering the short-range behavior 
of the one-body spin-density matrix.  Applications to several 
two-dimensional systems confirm
the excellent performance of the derived approximations, and 
verify the gauge-invariance requirement to be of great importance for dealing 
with current-carrying states. 
\end{abstract}


\maketitle

\section{Introduction}

The success of density-functional theory~\cite{dft} (DFT)
crucially depends on the availability of 
good approximations for the exchange-correlation (xc) energy functional.
Great progress has been achieved beyond
the commonly used local (spin) density 
approximation [L(S)DA] by means of, e.g., generalized-gradient
approximations, orbital functionals, and hybrid 
functionals~\cite{functionals}.
However, most of the approximations developed
so far have focused on three-dimensional (3D) systems. 

Most density functionals developed for 3D perform poorly when
applied to two-dimensional (2D) systems~\cite{kim,pollack,cost1,cost2}.
However, at present 2D structures constitute a large pool of 
applications in semiconductor nanotechnology, e.g., quantum Hall 
systems, spintronic devices, and quantum dots (QDs)~\cite{qd}.
Within DFT these systems are often treated using the 2D form of the LSDA
employing the analytic expression of exchange energy of the 2D electron
gas~\cite{rajagopal} (2DEG) and the corresponding correlation 
energies parametrized
via quantum Monte Carlo calculations first by Tanatar and Ceperley
\cite{tanatar} and later, for the complete range of collinear 
spin polarization, 
by Attaccalite {\em et al}~\cite{attaccalite}. Despite the relatively good
performance of the LSDA in terms of total energies, there is still a
clear need for accurate density functionals in 2D.
Only recently, xc functionals tailored for 2D have been pushed 
forward~\cite{pittalis1,nicole,correlation1,dga,ring,correlation2}. 
Here we make another natural step along these directions 
by studying Gaussian approximations (GAs) for the exchange energy functional. 
We extend previous studies on GAs~\cite{GBP,Kemi,Berko,LeeParr,PRG,Ayers}
to (i) 2D systems and to (ii) current-carrying states
for {\em both} 2D and 3D systems. 
Numerical tests of the resulting exchange energies
show significant improvement over the LSDA.

\section{General formalism}

Within current-spin-density functional theory~\cite{VR1,VR2} (CSDFT), the ground 
state energy $E_{v}$, the spin densities $\rho_{\sigma}(\vr)$ and paramagnetic
current density $\vj_{\sigma}(\vr)$ of a system 
of $N=N_{\uparrow}+N_{\downarrow}$ interacting electrons are determined
by minimizing the total-energy functional which is given in 
Hartree atomic units as
\begin{eqnarray}
&& E_{v}[\left\{ \rho_{\sigma} , \vj_{\sigma} \right\}] =  
T_s[\left\{ \rho_{\sigma} , \vj_{\sigma} \right\}]  + E_{\rm H}[\rho]\nonumber \\
& + & \sum_{\sigma=\uparrow,\downarrow}\int{d^{D}r} \; v_{\sigma}(\vr)
\rho_{\sigma}(\vr) + \frac{1}{c} \sum_{\sigma=\uparrow,\downarrow}\int{d^{D}r} \; {\bf A}_{\sigma}(\vr)
\vj_{\sigma}(\vr) \nonumber \\
&+&  \frac{1}{2c} \sum_{\sigma=\uparrow,\downarrow}\int{d^{D}r} \; {\bf A}^2_{\sigma}(\vr)
\rho_{\sigma}(\vr) 
 + E_{xc}[\left\{ \rho_{\sigma} \right\},\left\{ \vj_{\sigma} \right\}]
\label{etot}
\end{eqnarray}
where $D=2,3$ is the dimensionality of interest, 
$T_s[\left\{ \rho_{\sigma}, \vj_{\sigma} \right\}]$ is the kinetic energy functional of 
non-interacting particles, $v_{\sigma}(\vr)$ is an external (local) spin-dependent potential acting 
upon the interacting system, ${\bf A}_{\sigma}(\vr)$ is  an external (local) spin-dependent vector potential
(note that physical fields are spin-independent, but it is formally convenient to
introduce the spin-dependence), $E_{\rm H}[\rho]$ 
is the classical electrostatic or Hartree energy of the total charge density 
$\rho(\vr)=\rho_{\uparrow}(\vr)+\rho_{\downarrow}(\vr)$, and  
$E_{xc}[\left\{ \rho_{\sigma}, \vj_{\sigma} \right\}]$ is the xc 
energy functional. This quantity is usually further decomposed into
the exchange and correlation energies as
\begin{equation}
E_{xc}[\left\{ \rho_{\sigma} , \vj_{\sigma} \right\}]=
E_{x}[\left\{ \rho_{\sigma} ,\vj_{\sigma} \right\}]+E_{c}[\left\{ \rho_{\sigma} , \vj_{\sigma} \right\}]\;,
\end{equation}
An exact expression is known for the exchange energy 
\begin{equation}
E_x[\left\{ \rho_{\sigma} , \vj_{\sigma} \right\}] =
- \frac{1}{2} \sum_{\sigma=\uparrow,\downarrow} 
\int d^{D}r_1 \int d^{D}r_2 \frac{\big|\gamma_{\sigma}(\vr_1,\vr_2)\big|^2}{|\vr_1-\vr_2|}\;,
\label{EX_2}
\end{equation}
where $\gamma_{\sigma}$ is the one-body spin-density matrix (1BSDM) of
the Kohn-Sham (KS) system
\begin{equation}
\gamma_{\sigma}(\vr_1,\vr_2) =
\sum_{k=1}^{N_\sigma}\psi_{k,\sigma}(\vr_1)\psi^*_{k,\sigma}(\vr_2)\;,
\label{gamma}
\end{equation}
and $\psi_{k\sigma}(\vr)$ are the KS spin orbitals. Note that
we are assuming that the KS ground state has the form of a single Slater determinant. 
The number of electrons with spin $\sigma$ is given by
\begin{equation}
N_{\sigma} = \int d^Dr_1 \int d^Dr' \big|\gamma_{\sigma}(\vr_1,\vr')\big|^2\;.
\label{norm3}
\end{equation}
Introducing the average and relative particle coordinates,
respectively, as 
\begin{equation}
\vr = \frac{1}{2}\left( \vr_1 + \vr_2 \right),~\vs = \vr_1 - \vr_2\;,
\label{acic}
\end{equation}
Eq.~(\ref{EX_2}) can be rewritten as
\begin{equation}
E_x [\rho_{\uparrow},\rho_{\downarrow}] 
= - \frac{\Omega_D}{2} \sum_{\sigma=\uparrow,\downarrow} 
\int d^Dr \int s^{(D-2)}\,ds \big<\big|\gamma_{\sigma}\big|^2\big>,
\label{EX_3}
\end{equation}
where 
\begin{equation}
\big<\big|\gamma_{\sigma}\big|^2\big> = 
\frac{1}{\Omega_D} \int d\Omega_D
\Big|\gamma_{\sigma}\left(\vr+\frac{\vs}{2},\vr - \frac{\vs}{2}\right)\Big|^2 
\label{AG_1}
\end{equation}
is the angular average of $|\gamma_{\sigma}|^2$. 
Note that $\big<\big|\gamma_{\sigma}\big|^2\big>$ is a function of
the position, $\vr$, but this fact is omitted in the notation. 
Also in the remaining part of the present
work, we often shorten the notation where the dependence on $\vr$ is clear.
The term $d\Omega_{D}$ in Eq. (\ref{AG_1})
indicates the angular differential in $D$ dimensions, 
$\Omega_{2}=2\pi$, and $\Omega_{3}=4\pi$. 
In the new coordinates, Eq.~(\ref{norm3}) can be rewritten as
\begin{equation}
N_{\sigma} = \Omega_D \int d^Dr \int s^{(D-1)}ds \, \big<\big|\gamma_{\sigma}\big|^2\big>\,.
\label{norm4}
\end{equation}
It is apparent that an approximation for
$\big<\big|\gamma_{\sigma}\big|^2\big>$ 
in Eq. (\ref{AG_1}) provides an approximation for 
the exchange energy in Eq. (\ref{EX_3}).
Equation (\ref{norm4}) allows us to check the validity of such an approximation 
in terms of the particle-number normalization.

\section{Gaussian approximation for the one-body spin-density matrix}

One of the more successful strategies to develop approximation for the
exchange energy functional is based on the density-matrix expansion scheme.~\cite{Negele,Koehl,Voorhis,Tsuneda}
Here we reconsider the so-called short-range behavior of the 1BSDM. 
In particular, we
extend the results previously derived for 
zero-current states in 3D~\cite{Kemi,Berko} to 
{\em current-carrying} states of {\em both} 3D and 2D systems. 

In dealing with current-carrying states 
it is necessary to allow the orbitals in Eq.~(\ref{gamma})
to be complex valued. Expanding the 1BSDM in the inter-particle 
coordinate, $\vs$, at a given average position, $\vr$, we find
\begin{equation}
\gamma_{\sigma}\left(\vr+\frac{\vs}{2},\vr - \frac{\vs}{2}\right) \approx 
\rho_{\sigma}(\vr)\left\{1 - \frac{\vs^{T}\bf{\Sigma}_{\sigma}\vs}{\rho_{\sigma}} 
+ i \frac{\vj^{T}_{\sigma}\vs}{\rho_{\sigma}} \right\}\;,
\label{taylor}
\end{equation}
where
\begin{equation}
\vj_{\sigma}=\frac{1}{2i}\sum_{k=1}^{N_\sigma} \left[
\psi^*_{k\sigma} \left(\nabla \psi_{k\sigma}\right) 
- \left(\nabla \psi^*_{k\sigma}\right) \psi_{k\sigma} \right]
\end{equation}
is the spin-dependent (paramagnetic) current density, and
\begin{equation}\label{SIGM}
\left[ \bf{\Sigma}_{\sigma} \right]_{\alpha,\beta} = \left[ t_{\sigma} \right]_{\alpha,\beta} 
+ \frac{1}{2} \sum_{k=1}^{N_\sigma} \frac{j_{k\sigma,\alpha}j_{k\sigma,\beta}}{\rho_{k\sigma}}\,,
\end{equation}
with
\begin{equation}
\left[ t_{\sigma} \right]_{\alpha,\beta} = \frac{1}{8} \left[ \sum_{k=1}^{N_\sigma}
\frac{1}{\rho_{k\sigma}}
\frac{\partial \rho_{k\sigma}}{\partial x_{\alpha}} \frac{\partial \rho_{k\sigma} }{ \partial x_{\beta}}
- \frac{\partial^2}{\partial x_{\alpha}\partial x_{\beta}}\rho_{\sigma} \right]\,,
\end{equation}
where $\alpha,\beta$ runs over the Cartesian coordinates in 3D or 2D. 
The term $\left[ \bf{\Sigma}_{\sigma} \right]_{\alpha,\beta}$
can be interpreted as kinetic energy-density tensor. We point out that
$\bf{\Sigma}_{\sigma}$ is not a gauge-invariant quantity 
due to the second term in Eq. (\ref{SIGM}).

At this point, we carry out an exponential resummation 
of Eq.~(\ref{taylor}), which yields
\begin{equation}
\tilde{\gamma}_{\sigma} = 
\rho_{\sigma} 
\exp{\left[ 
- \frac{\vs^{T}\bf{\Sigma}_{\sigma}\vs}{\rho_{\sigma}} 
+ i \frac{\vj^{T}_{\sigma}\vs}{\rho_{\sigma}} \right]}\;,
\label{AG_w}
\end{equation}
where $\vs^{T}$ and $\vj^{T}$ are the transposes of the vectors $\vs$ and $\vj$, respectively.
Although this approximation reproduces 
the exact short range behavior of the 1BSDM,  
it seems -- at first sight -- a rather oversimplified approximation for arbitrary $\vs$. 
However,  it has the appealing feature that the corresponding Wigner transform \cite{dft} 
reproduces the spin-particle density, the paramagnetic current
density, and the kinetic-energy-density 
tensor of the KS system exactly (see below). 

The Wigner transformation of Eq. (\ref{AG_w}) is given by
{\small
\begin{eqnarray}
\!\!\!\!\!\!\tilde{f}^W_{\sigma}(\vr,\vp) \!\!\! &\approx& \!\!\!
\frac{\rho_{\sigma}}{\sqrt{(2\pi)^D~{\mathrm det}~ \left( k_B {\boldsymbol \Theta}_{\sigma} \right)}} \nonumber \\
\!\!\!\!\!\!\!\!\! &\times& \!\!\! \exp{\left[
-\frac{1}{2}\left(\vp-\frac{\vj_{\sigma}}{\rho_{\sigma}}\right)^{T}(k_B\,{\boldsymbol \Theta}_{\sigma})^{-1}
\left(\vp-\frac{\vj_{\sigma}}{\rho_{\sigma}}\right)\right]}\!\!\;,
\label{AW}
\end{eqnarray}
}
where, following Ref.~\cite{Berko}, we 
define the matrix ${\boldsymbol \Theta}_{\sigma}$ to satisfy the relation
\begin{equation}
\frac{1}{2} k_B \, \rho_{\sigma} {\boldsymbol \Theta}_{\sigma} = \bf{\Sigma}_{\sigma},
\label{idgas}
\end{equation}
where $k_B$ is the Boltzmann constant. The presence of the Boltzmann constant
is motivated by the fact that for the spin-unpolarized zero-current case, 
the Wigner transform in question provides a description
for the KS state formally resembling a thermodynamic one~\cite{GBP,Kemi,Berko,LeeParr,PRG,Ayers}.
Instead, in this work 
we allow the orbitals to be spin-unrestricted and complex valued. 
We observer that ${\boldsymbol \Theta}_{\sigma}$ is apparently not gauge-invariant, thus we shall
refrain from interpreting it as a local temperature.
Finally, it is immediate to verify that 
\begin{equation}
\rho_{\sigma}(\vr) = \int d^D p  \, \tilde{f}^W_{\sigma}(\vr,\vp)\,,
\end{equation}
\begin{equation}
\vj_{\sigma}(\vr) =  \int d^D p  \, \vp  \, \tilde{f}^W_{\sigma}(\vr,\vp)\,,
\end{equation}
and
\begin{eqnarray}
\left[ \bf{\Sigma}_{\sigma} \right]_{\alpha,\beta}(\vr) & = & \frac{1}{2} \int d^D p   \,
\left( p_{\alpha}-\frac{j_{\sigma,\alpha}(\vr)}{\rho_{\sigma}(\vr)}\right) 
\left( p_{\beta} -\frac{j_{\sigma,\beta}(\vr)}{\rho_{\sigma}(\vr)} \right) \nonumber \\
& \times & \tilde{f}^W_{\sigma}(\vr,\vp)\,.
\end{eqnarray}

\section{Gaussian approximations for the exchange energies}

In this section, we provide few approximations for the
exchange energy of spin-polarized current-carrying states. 
Although what follows may be applied to the 3D case as well, 
we shall restrict ourselves to 2D systems where the need of 
new approximations in the quantum-Hall regime, for example,
is particularly large. 

Let us first observe that substitution of Eq. (\ref{AG_w}) in Eq. (\ref{AG_1}) 
would yield an expression which is not gauge invariant. 
Popular (3D) meta-generalized gradient approximations 
for the xc energy are also not gauge invariant~\cite{JTao2}.
This fact is not only formally inadequate~\cite{VR1,VR2}, but 
it is also a source of practical problems.
For example, it has been observed that the lack of gauge invariance causes
a wrong description of degenerate atomic ground states carrying 
different paramagnetic currents. Recently, Tao and Perdew have 
proposed a correction for this problem~\cite{JTao,JTao2}. 
Here we show how this kind of correction naturally emerges in the 
context of the present GA.

In order to fulfill the gauge invariance requirement, we 
substitute Eq. (\ref{taylor}) in Eq. (\ref{AG_1}), and obtain 
\begin{equation}
\big<\big|\gamma_{\sigma}\big|^2\big>  \approx 
\rho_{\sigma}^2 \left\{ 1 -\left[ \frac{{\mathrm Tr}\left(\bf{\Sigma}_{\sigma}\right)}{\rho_\sigma}
-\frac{1}{2}\left(\frac{\vj_{\sigma}}{\rho_{\sigma}}\right)^2
\right]s^2 \right\}
\label{AG_4}
\end{equation}
where
\begin{eqnarray}
 {\mathrm Tr}\left(\bf{\Sigma}_{\sigma}\right) &=& 
\sum_{k=1}^{N_\sigma}
\left[\frac{1}{8} \frac{\left(\nabla \rho_{k\sigma}\right)^2}{\rho_{k\sigma}} 
+ \frac{1}{2}\frac{\vj^2_{k\sigma}}{\rho_{k\sigma}}\right]
- \frac{1}{8}\nabla^2\rho_{\sigma}\nonumber \\
&=&  \tau_{\sigma}-\frac{1}{8}\nabla^2\rho_{\sigma}\;,
\label{trace}
\end{eqnarray}
with
\begin{equation}
\tau_{\sigma}=\frac{1}{2}\sum_{k=1}^{N_\sigma} |\nabla\psi_{k,\sigma}|^2
\end{equation}
being the spin-dependent kinetic-energy density. Note that, while
$\tau_{\sigma}$ is not gauge invariant, the modified quantity
\begin{equation}
\tilde{\tau}_{\sigma}=\tau_{\sigma} - \frac{1}{2}\left(\frac{\vj^2_{\sigma}}{\rho_{\sigma}}\right)
\label{tau2}
\end{equation}
is indeed gauge invariant.
Now, let us perform an exponential resummation of Eq. (\ref{AG_4}), which leads to the
following GA
\begin{equation}
\big<|\gamma_{\sigma}|^2\big>  \approx 
\rho_{\sigma}^2 \exp{\left(-\frac{s^2}{\beta_{\sigma}}\right)},
\label{AG_5}
\end{equation}
where
\begin{equation}
\beta_{\sigma}^{-1} = \left[\frac{\tau_{\sigma}}{\rho_\sigma} 
- \frac{1}{8} \frac{\nabla^2\rho_{\sigma}}{\rho_\sigma}
-\frac{1}{2}\left(\frac{\vj_{\sigma}}{\rho_{\sigma}}\right)^2
\right],
\label{temp1}
\end{equation}
or, alternatively, using Eq. (\ref{idgas})
\begin{equation}
\beta_{\sigma}^{-1} =  \frac{1}{2} \left[
{\mathrm Tr}\left({\boldsymbol \Theta}_\sigma\right) - \left(\frac{\vj_{\sigma}}{\rho_{\sigma}}\right)^2 \right] \,.
\label{temp2}
\end{equation}
Expression (\ref{temp2}) shows how $\beta_{\sigma}$ relates to 
${\boldsymbol \Theta}_\sigma$. In contrast to ${\boldsymbol \Theta}_\sigma$,
$\beta_{\sigma}$ is a gauge-invariant quantity. Hence, the 
approximation for the exchange energy obtained from Eq. (\ref{temp2}) is gauge invariant as well.

We further focus on inhomogeneous 2D systems. 
An expression for the exchange energy is readily obtained
by inserting Eq. (\ref{AG_5}) in Eqs. (\ref{EX_3}) and (\ref{norm4}):
\begin{equation}
E_{x} [\rho_{\uparrow},\rho_{\downarrow}] 
= - \frac{\pi^{3/2}}{2} \sum_{\sigma=\uparrow,\downarrow} 
\int d^{2}r \rho^2_{\sigma}(\vr)\beta^{{1/2}}_{\sigma}(\vr)\,.
\label{AEX_3}
\end{equation}
This is a functional for the exchange 
energy in the form of a current-dependent meta-generalized 
gradient approximation.
To stress the dependency on the paramagnetic current, and to remind its 
origin from a GA, we will indicate the functional as J-GA.

Furthermore, we impose that our approximation reproduces 
the correct normalization
condition (\ref{norm3}). Hence, we modify Eq.~(\ref{AG_5}) by adding 
a fourth-order term~\cite{LeeParr} as follows 
\begin{equation}
\big<|\gamma_{\sigma}|^2\big>  \approx 
\rho_{\sigma}^2 \exp{\left(-\frac{s^2}{\beta_{\sigma}}\right)} \left[ 1 + A_\sigma \left(  \frac{s}{\beta_{\sigma}} \right)^2
\right]\,.
\label{AG_6}
\end{equation}
This expression leads to 
\begin{eqnarray}
E_{x} [\rho_{\uparrow},\rho_{\downarrow}] 
& = & - \frac{\sqrt{\pi}}{2} \sum_{\sigma=\uparrow,\downarrow} 
\int d^{2}r \left[ \pi + \frac{3}{4} \sqrt{\pi} A_{\sigma} \right]
\nonumber \\ 
& \times & \rho^2_{\sigma}(\vr)\beta^{1/2}_{\sigma}(\vr)\,.
\label{AEX_4}
\end{eqnarray}
The parameter $A_\sigma$ is determined by using the following relation
\begin{equation}
N_{\sigma} = \pi \int d^{2}r \left[ 1 + 2 A_{\sigma} \right] \rho^2_{\sigma}(\vr) \beta_{\sigma}(\vr)\,.
\label{Anorm4}
\end{equation}
Eq. (\ref{AEX_4}) together with Eq. (\ref{Anorm4}) provide another
density functional for the exchange energy. We refer to this
functional as the current-dependent {\em modified} GA denoted by 
J-MGA.

\section{Applications}

In order to test the performance and the degree of universality 
of the J-GA and J-MGA, given in Eqs.~(\ref{AEX_3}) and 
(\ref{AEX_4}), respectively, we consider below different 2D electron 
systems, i.e., QDs of different shapes, as well as the 2DEG. 
We also assess the importance of the approximation to account 
for the paramagnetic current density in Eqs.~(\ref{AG_4}) 
and (\ref{trace}), $\vj_\sigma$, and thus the prerequisite of the gauge
invariance. The assessment is done by comparing J-GA and J-MGA to
the results obtained by {\em neglecting} in the functionals 
the terms depending explicitly on  $\vj_\sigma$. These approximations
are denoted below as 0-GA and 0-MGA. 

\subsection{Finite systems}\label{dots}

As relevant examples of finite 2D electron systems we consider
parabolic (harmonic)~\cite{qd} and rectangular QDs~\cite{recta_qd,recta_qd2}, 
respectively.
In both cases, the many-electron Hamiltonian is given by
\begin{equation}
H = \sum^N_{i=1}\left[\frac{1}{2}\left({\mathbf p}_i+
\frac{1}{c}{\bf A}_i\right)^2
+V_{\rm ext}(\vr_i) \right]
+ \sum^N_{i<j}\frac{1}{|{\mathbf r}_i-{\mathbf r}_j|},
\label{hamiltonian}
\end{equation}
where $N$ is the number of electrons and
${\bf A}$ is the external vector potential 
of the homogeneous magnetic field 
${\bf B}=B{\hat z}$ perpendicular to the 2D ({\em x-y}) plane.

The external confining potential is defined for a parabolic
QD, containing here $N=2\ldots 20$ electrons, as 
\be
V_{\rm ext}^{\rm par}(r)=\frac{1}{2}\omega_0^2 r^2,
\ee
where $\omega_0$ is the confinement strength.
For $N=2$ we have $\omega_0=1$ (see below), and otherwise
$\omega_0=0.42168$ corresponding to a typical confinement of $5$ meV
when applying the effective mass approximation (with the effective mass
$m^*=0.067\,m_e$ and dielectric constant $\epsilon=12.4\,\epsilon_0$)
in the modeling of GaAs QDs~\cite{qd}.

The rectangular QD containing $N=4\ldots 16$ electrons 
is defined by
\be
V_{\rm ext}^{\rm rec}(x,y)=\left\{ \begin{array}{ll}
0, & 0\leq{x}\leq\alpha{L},\,0\leq{y}\leq{L}\\
\infty, & \text{ elsewhere},
\end{array} \right.
\ee
where $L=\sqrt{2}\pi$ is the (smaller) side length, and
$\alpha=2$ determines the side ratio of the rectangle. The
size is then $\sim 90\,{\rm nm}\times 45\,{\rm nm}$. The electronic properties of
similar rectangular QDs have been studied in detail in 
Refs.~\cite{recta_qd} and \cite{recta_qd2}.

We point out that, in principle, the presence of ${\bf A}$
in the Hamiltonian brings out the need of the
full-flagged current-spin-density functional theory~\cite{VR1}. 
However, at the level of the
KS equations we neglect the xc vector potential, but
still include the external one.
This approximation
has been shown to be accurate for atomistic systems, and
QDs up to relatively high magnetic
fields~\cite{nicole,pittalis1,pittalis2,pittalis3,Handy,henri,capelle1,capelle2}.

As reference we use the exact-exchange (EXX)
results calculated in the Krieger-Li-Iafrate (KLI) 
approximation~\cite{kli}.
The converged KS orbitals from the KLI are then
used as the input for our functionals. 
An exception is the two-electron parabolic QD with 
$\omega_0=1$, for which we can apply the known 
{\em analytic} density~\cite{taut} in the calculation of the 
EXX energy (which is minus half of the Hartree energy)
and as the input in the functionals.
For comparison, we also
compute the exchange energies from the (2D) LSDA~\cite{rajagopal}.
Both the EXX and LSDA calculations are performed using the 
{\tt octopus} real-space code 
within SDFT~\cite{octopus}.

Table~\ref{table_closed}
\begin{table}
\caption{\label{table_closed}
Exchange energies for closed-shell parabolic (upper part) and
rectangular (lower part) quantum dots
calculated using the exact exchange (EXX) (in the KLI approximation
for $N>2$), the Gaussian approximation (GA), 
the modified Gaussian approximation (MGA), and
the local spin-density approximation (LSDA).
The relative errors with respect to EXX are also shown.
}
\begin{tabular*}{\columnwidth}{@{\extracolsep{\fill}} c c | c c | c c | c c}
\hline
\hline
$N$ & EXX & J-GA & $\Delta$ (\%) & J-MGA & $\Delta$ (\%) & LSDA & $\Delta$ (\%)\\
\hline
2  & -1.08 & -1.12 & -3.0 & -1.10 & -1.5 & -0.98 & 9.3 \\
6  & -2.23 & -2.28 & -2.1 & -2.28 & -2.2 & -2.13 & 4.4 \\
12 & -4.89 & -5.01 & -2.5 & -5.03 & -2.9 & -4.76 & 2.6 \\
20 & -8.78 & -9.00 & -2.5 & -9.05 & -3.0 & -8.63 & 1.7 \\
\hline
6  & -3.14 & -3.33 & -5.9 & -3.25 & -3.3  &2.99 &  4.9\\
12 & -8.19 & -8.46 & -3.3 & -8.42 & -2.8  &-7.99 & 2.5 \\
16 & -12.7 & -13.3 & -4.4 & -13.1 & -3.2  &-12.3 & 3.5 \\
\hline
\end{tabular*}
\end{table}
shows the exchange energies for a set of both parabolic (upper part) and 
rectangular (lower part) QDs. 
The relative errors with respect to EXX are also shown.
All the cases correspond to closed-shell
ground-state solutions possessing no currents, and the total spin is $S=0$.
Therefore, the J-GA (J-MGA) result is similar to 0-GA (0-MGA).
Overall, we find a good agreement between the J-GA and the EXX. 
The J-MGA, which contains the fourth-order
term in $\big<|\gamma_\sigma|^2\big>$, is in most examples 
more accurate: The deviation from the EXX is $\lesssim 3\,\%$.
As expected, the LSDA generally overestimates
the exchange energies, but the deviation from the EXX reduces 
as a function of $N$. 
In contrast, J-(M)GA seems to gradually lose its accuracy
when $N$ is increased, especially in parabolic QDs. 
This may be due to the fact that since the electron density profile is 
relatively flat in large QDs, the LSDA correspondingly becomes more
suitable.

In Table~\ref{table_currents}
\begin{table*}
\caption{\label{table_currents}
Exchange energies (in effective atomic units, see Sec.~\ref{dots}) 
for fully spin-polarized ($S=N/2$), current-carrying 
states in parabolic (upper part) and rectangular (lower part) quantum dots.
The results have been calculated using the exact exchange (EXX) in the
KLI approximation, 
the current-dependent Gaussian approximation (J-GA), 
the current-dependent modified GA (J-MGA), 
the non-current-dependent GA (0-GA) and MGA (0-MGA), and
the local spin-density approximation (LSDA).
The relative errors with respect to EXX are also shown.
}
\begin{tabular}{ c c c | c c | c c | c c | c c | c c }
\hline
\hline
$N$ & $B$ (T) & EXX & J-GA & $\Delta$ (\%) & J-MGA & $\Delta$ (\%) & 0-GA & $\Delta$ (\%) & 0-MGA & $\Delta$ (\%) & LSDA & $\Delta$ (\%)\\
\hline
2 & 0 & -1.30 & -1.26 & 3.0 & -1.29 & 1.2  &-1.17 & 9.8 & -1.33 & -2.3 &-1.21 & 6.8\\
4 & 4 & -1.92 & -1.89 & 1.8 & -1.92 & 0.21 &-1.49 & 23 & -2.18  & -13 &-1.83 & 4.8\\
6 & 6 & -3.26 & -3.21 & 1.5 & -3.26 & 0    &-2.26 & 31 &-3.99  &  -22 &-3.11 & 4.5\\
\hline
4  & 8  & -2.46 & -2.43 & 1.3  & -2.46 & 0 & -1.66 & 32 &-2.07 & 16 & -2.36 & 4.3\\
6  & 10 & -4.30 & -4.26 & 1.0  & -4.31 & -0.19 &-2.71 & 37 &-3.51 &18 &-4.14 & 3.9\\
8  & 10 & -6.42 & -6.42 &  0   & -6.46 & -0.67 &-4.00 & 38 &-5.24 & 18&-6.18 & 3.8\\
12 & 12 & -11.6 & -11.6 &  0   & -11.7 & -0.86 &-6.67 & 42 &-9.30 & 20&-11.1 & 3.6\\
\hline
\end{tabular}
\end{table*}
we show the exchange energies for spin-polarized current-carrying
states in parabolic (upper part) and rectangular (lower part) QDs.
The $N=2$ QD with $S=1$ corresponds to an excited state, whereas
in the other systems the spin polarization of the ground state 
is achieved by applying the external magnetic field. 
Compared to the EXX, the accuracy of both J-GA and J-MGA is
excellent. Moreover, both functionals are considerably more 
accurate than the LSDA. On the other hand, the approximations 
0-GA and 0-MGA obtained by omitting the current-dependent 
terms in Eq.~(\ref{AG_4})
lead to completely wrong results. This demonstrates the importance
of the currents involved in the derivation of a gauge invariant GA.

\subsection{Two-dimensional electron gas}

Finally let us consider the GA for the exchange energy in the case of 
the 2DEG. We restrict ourselves to the case of vanishing external 
vector potential, but still we allow the external scalar potential 
to be spin-dependent.
In this situation, we can set $\nabla^2\rho_{\sigma}=0$ and 
$\nabla\rho_{\sigma}=0$.
Moreover, we may set $\vj_{\sigma}=0$ by considering 
the 2DEG at rest. Hence, the current term in Eq. (\ref{temp1}) 
drops out from the final expression
\begin{equation}
\big<|\gamma_{\sigma}|^2\big>  \approx 
\rho_{\sigma}^2 \exp{\left(-\pi\rho^2_{\sigma}s^2\right)}.
\label{GLDA}
\end{equation}
We note that correct particle numbers for each spin channel are 
obtained from Eq. (\ref{GLDA}), 
and therefore $A_\sigma = 0$.
Defining the 2D density parameter $r_s=1/\sqrt{\pi\rho}$, and
the polarization
$\xi=(\rho_{\uparrow}-\rho_{\downarrow})/\rho$, the total exchange
energy per particle becomes
\begin{equation}
\epsilon_{x}[r_s,\xi]=-\frac{\sqrt{\pi}}{4\sqrt{2}\,r_s}\left[(1+\xi)^{3/2}+(1-\xi)^{3/2}\right].
\label{final_xenergy_N}
\end{equation}
This expression can be used as an explicit 
density functional in the LSDA fashion. In fact, the only difference
to the LSDA exchange~\cite{rajagopal} is the prefactor, which is here 
$\sim 4.4\%$ smaller than in the LSDA. Therefore, one can immediately  
deduce from Tables~\ref{table_closed} and \ref{table_currents} that, 
when considering finite QD systems of few electrons, 
Eq.~(\ref{final_xenergy_N}) 
leads to good agreement with the EXX results by correcting the LSDA 
exchange energies by $\sim 4.4\%$. 

Interestingly, Eq.~(\ref{final_xenergy_N}) has been
recently obtained in an alternative way
by considering the short-range behavior 
of the exchange-hole function~\cite{pittalis1}. 
This identity of the expression can be seen as a rather
reassuring result.

\section{Conclusions}
 
In this work we have analyzed the one-body spin-density matrix for
current-carrying states.
We devised gauge-invariant approximations for the exchange energy
by considering a Gaussian approximation for the short-range behavior 
of the one-body spin-density matrix, in particular, for its 
angular average of the module square.
The resulting simple approximations for the exchange energy
perform extremely well in a variety of two-dimensional systems, 
including a diverse set of quantum dots, as well as the homogeneous 
two-dimensional electron gas. Moreover, our analysis confirms the 
relevance of the gauge invariance when dealing with current-carrying 
states.

\begin{acknowledgments}
This work was supported by the Deutsche 
Forschungsgemeinschaft, the EU's Sixth Framework
Programme through the ETSF e-I3, and the Academy of
Finland.
\end{acknowledgments}

\end{document}